\documentclass[apj, twocolumn]{emulateapj}

\usepackage{apjfonts}
\usepackage{amsmath}
\usepackage{longtable}
\usepackage{epsfig}
\usepackage{hhline}

\begin{document}

\title{On the discovery of the first 350 micron-selected galaxy}

\author{Sophia~A.~Khan\altaffilmark{1,2}, Richard~A.~Shafer\altaffilmark{2}, Dominic~J.~Benford\altaffilmark{2}, Johannes~G.~Staguhn\altaffilmark{2,3}, Pierre~Chanial\altaffilmark{2,4}, Emeric~Le~Floc'h\altaffilmark{5}, Thomas~S.~R.~Babbedge\altaffilmark{1},
Duncan~Farrah\altaffilmark{6,7}, S.~Harvey~Moseley\altaffilmark{2}, Eli~Dwek\altaffilmark{2}, David~L.~Clements\altaffilmark{1}, Timothy~J.~Sumner\altaffilmark{1}, Matthew~L.~N.~Ashby\altaffilmark{8}, Kate~Brand\altaffilmark{9}, Mark~Brodwin\altaffilmark{10}, Peter~R.~Eisenhardt\altaffilmark{10}, Richard~Elston\altaffilmark{11,12}, Anthony~H.~Gonzalez\altaffilmark{11}, Eric~McKenzie\altaffilmark{11}, Stephen~S.~Murray\altaffilmark{8}}

\altaffiltext{1}{Imperial College London, Blackett Laboratory, Prince 
Consort Road, London SW7 2AZ, UK}
\altaffiltext{2}{Observational Cosmology Laboratory (Code 665), NASA 
/ Goddard Space Flight Center, Greenbelt, MD 20771}
\altaffiltext{3}{SSAI, Lanham, MD 20706}
\altaffiltext{4}{NRC}
\altaffiltext{5}{Steward Observatory, University of Arizona, 933 N. 
Cherry Avenue, Tucson, AZ 85721}
\altaffiltext{6}{Spitzer Science Center, MC 220-6, 1200 East 
California Boulevard, Pasadena, CA 91125}
\altaffiltext{7}{Department of Astronomy, Cornell University, 610 Space Sciences Building, Ithaca, NY 14853-6801}
\altaffiltext{8}{Harvard Smithsonian Center for Astrophysics, 60 
Garden Street MS-66, Cambridge, MA 02138}
\altaffiltext{9}{National Optical Astronomy Observatory, Tucson, AZ 
85719-6732}
\altaffiltext{10}{Jet Propulsion Laboratory, California Institute of 
Technology, MC 169-327, 4800 Oak Grove Drive, Pasadena, CA 91109}
\altaffiltext{11}{Department of Astronomy, University of Florida, Gainesville, FL 32611}
\altaffiltext{12}{Deceased}

\begin{abstract}
We report the detection of a 3.6$\sigma$ 350\,$\mu$m-selected source in the
Bo\"{o}tes Deep Field.  The source, the first Short-wavelength
Submillimeter-selected Galaxy (SSG~1), was discovered as part of a
blank field extragalactic survey using the 350\,$\mu$m-optimised
Submillimeter High Angular Resolution Camera (SHARC~II) at the Caltech
Submillimeter Observatory. With multiwavelength photometry from
NOAO-NDWFS (R and I band), FLAMEX (J and K$_{s}$), {\it Spitzer} (IRAC
and MIPS) and the Westerbork 1.4\,GHz Deep Survey (radio upper limit), we are able to 
constrain the photometric redshift using different methods, all of which suggest a redshift $\sim$1.  In the
absence of long-wavelength submillimeter data we use SED templates to
infer that this source is an ultraluminous infrared galaxy (ULIRG) 
with a dust temperature of 30$\pm$5\,K, occupying a region of luminosity-temperature space shared by modarate redshift {\it ISO}-selected ULIRGs (rather than high redshift SCUBA-selected SMGs).  SHARC~II can thus select SMGs with moderately ``warm'' dust that might be missed in submillimeter surveys at longer wavelengths.  
\end{abstract}


\section{INTRODUCTION}

Submillimeter-selected galaxies (SMGs) generally refer to the
population detected in the pioneering lensed and blank deep field
surveys using the Submillimeter Common User Bolometer Array (SCUBA; Holland et al. 1999)
instrument on the James Clerk Maxwell Telescope (JCMT) (e.g., Smail,
Ivison \& Blain 1997; Hughes et al.  1998; Barger et al.  1998; Eales
et al.  1999).  They are regarded as the high redshift (z$\sim$2--3; Chapman 
et al.  2005; Chapman et al. 2003; see also Simpson et al. 2004) counterparts to
the population of luminous and ultraluminous infrared galaxies (LIRGs
and ULIRGs) detected by the InfraRed Astronomical Satellite, {\it IRAS}, typically out
to z$\lesssim$0.1 (Soifer et al. 1984; Joseph \& Wright 1985; and Soifer,
Neugebauer \& Houck 1987), since they share similar properties (fundamentally, that the bulk of the bolometric luminosity is emitted in the restframe far-IR, 
powered by a combination of star formation and AGN activity.).  

As a necessary consequence of the selection at longer submillimeter wavelengths, SMGs tend to have higher bolometric luminosities, and those with comparable luminosities to {\it IRAS}-selected ULIRGs tend to have cooler dust temperatures (Blain et al.  2004).  Between the two are the Infrared Space Observatory ({\it ISO})-selected ULIRGs (Aussel et al. 1999; Rowan-Robinson et al. 1999; Elbaz et al. 2002a), which occupy an intermediate redshift space (z$\sim$1), and have dust masses in-between the {\it IRAS} and SCUBA-selected ULIRGs (Chapman et al.  2002; Blain et al.  2004; this is also a region of parameter space that will be shared by {\it Spitzer}-selected galaxies, see, e.g., Yan et al. 2004).  

Sources selected at 350\,$\mu$m are expected to be predominantly LIRGs and ULIRGs at 1$<z<$3 (the K-correction changes from negative to positive with increasing redshift as the observing window moves towards and past the Wien peak; see, e.g., Guiderdoni et al. 1998; Khan, 2005, in preparation).  Galaxies at z$\sim$1 are likely to be the dominant source of the integrated far-IR background (Puget et al.  1996; Fixsen et al.  1998; Elbaz et al. 2002b); also, observations suggest that after a relatively steep rise from z=0 to $\sim$1, the cosmic star formation rate (CSFR) appears to flatten off between z=1 and $\sim$4, with evidence for a slow decline at higher redshift (Lilly et al.  1996; Madau et al.  1996; Connolly et al.  1997; Steidel et al. 1999; Gabasch et al.  2004).  Hence 350\,$\mu$m-selected sources are an important probe of the epoch of peak star formation rate in the universe.

In this Letter we report on the detection of a single source above 3$\sigma$ at 350\,$\mu$m.  It is, to the best of our knowledge, the first short-wavelength submillimeter-selected galaxy (200--500\,$\mu$m) and is denoted as SSG~1 in this Letter (but by standard convention, SMM J143206.65+341613.4).  SSG~1 is the first object discovered {\it purely} by its 350\,$\mu$m emission, through blank deep observations of the Bo\"{o}tes field in a program designed to search efficiently for new sources at 350\,$\mu$m (the SHARC~II Unbiased Deep Survey, SUDS; Khan et al., 2005, in preparation).  Verification of the detection is confirmed by
coincident sources subsequently identified in the {\it Spitzer} MIPS
24\,$\mu$m filter, all four {\it Spitzer} IRAC filters, and J, K$_{s}$, I and R
band, enabling us to place constraints on the photometric redshift,
luminosity and nature of this source.  This Letter presents these fluxes, plus important flux upper limits in other bands, to characterize this source, and with this understanding we consider the implications for other sources that may 
be discovered at 350\,$\mu$m. 

\section{OBSERVATION}

Eight hours of data were obtained using the Second Generation
Submillimeter High Angular Resolution Camera (SHARC~II) at the Caltech
Submillimeter Observatory (CSO) on Mauna Kea, Hawai'i, in January and March
2004.  SHARC~II is a 350\,$\mu$m-optimized camera (Dowell et al.
2003) built around a $12\times 32$ element close-packed bolometer
array (Moseley et al.  2004).  It achieves a point-source sensitivity
of $\rm \sim 1\,Jy~Hz^{-1/2}$ in good weather.  The 384 pixels of the
SHARC~II array image a region of around $1.0' \times 2.5'$ on the sky.
Its filled absorber array provides instantaneous imaging of the entire
field of view, sampled at roughly 2.5 pixels per nominal beam area.
The beam profile was measured on known compact sources, and was
verified to be within three per cent of the diffraction-limited
beamwidth of $8.5''$.  All observations were taken using the Dish
Surface Optimisation System (Leong et al. 2003), which corrects for
the primary mirror deformation as a function of zenith angle, to
improve the telescope efficiency and the pointing.

For these data the in-band zenith atmospheric opacity
($\tau_{350\,\mu\rm m}$) ranged from 1.0 to 1.4, corresponding to a
zenith transmission of around 30 per cent.  Our observations were
centred on the Bo\"{o}tes Deep Field (de Vries et al.  2002), at
position RA$=14^h32^m5.75^s$, Dec$=34^\circ16'47.5''$ (J2000).

The data were reduced using the standard CSO reduction software, CRUSH
(Kov\'acs, 2005, in preparation).  This software implements a 
self-consistent
least-squares algorithm to solve for the celestial emission, taking
into account instrumental and atmospheric contributions to the signal.

The skymap is calibrated with the flux and point spread function based on 
observations of Callisto taken throughout the observing period at similar elevations.  

\section{RESULTS}

\begin{table}
    \caption{Multiwavelength photometry for SSG~1}
   \label{tab:fluxes}
   \begin{center}\begin{tabular}{||c|c|c||}
	\hhline{|t:===:t|} & & \\ [-0.15 cm]
	Observed wavelength & Flux density   	       & 
Instrument--Survey       \\ [0.1 cm]
	\hhline{||---||}
	\hhline{||---||} & & \\ [-0.15 cm]
	X-ray (0.5--7\,keV)&$< 4\times 10^{-15}$c.g.s& {\it Chandra}--XBo\"{o}tes 
\\
	4220\AA             & $<$ 0.1\,$\mu$Jy        & KPNO--NDWFS 
\\
	6590\AA             & 1.0  $\pm$ 0.1\,$\mu$Jy  & KPNO--NDWFS 
\\
	8081\AA             & 3.1  $\pm$ 0.2\,$\mu$Jy  & KPNO--NDWFS 
\\
	1.24\,$\mu$m        & 10.5 $\pm$ 1.1\,$\mu$Jy  & KPNO--FLAMEX 
\\
	2.16\,$\mu$m        & 42.0 $\pm$ 1.4\,$\mu$Jy  & KPNO--FLAMEX 
\\
	3.6\,$\mu$m         & 80.0 $\pm$ 2.9\,$\mu$Jy  & {\it Spitzer} 
IRAC--Shallow   \\
  	4.5\,$\mu$m         & 65.8 $\pm$ 3.6\,$\mu$Jy  & {\it Spitzer} 
IRAC--Shallow   \\
	5.8\,$\mu$m         & 55.3 $\pm$ 17.3\,$\mu$Jy & {\it Spitzer} 
IRAC--Shallow   \\
	8.0\,$\mu$m         & 81.7 $\pm$ 15.1\,$\mu$Jy & {\it Spitzer} 
IRAC--Shallow   \\
	24\,$\mu$m          & 0.61 $\pm$ 0.04\,mJy     & {\it Spitzer} 
MIPS--IRS/MIPS GTO    \\
	70\,$\mu$m          & $<$ 40\,mJy             & {\it Spitzer} 
MIPS--IRS/MIPS GTO      \\
	160\,$\mu$m         & $<$ 100\,mJy            & {\it Spitzer} 
MIPS--IRS/MIPS GTO     \\
	350\,$\mu$m         & 23.2 $\pm$ 7.9\,mJy     & SHARC~II--SUDS     \\
	21\,cm              & $<$ 63\,$\mu$Jy         & Westerbork--Deep 
1.4\,GHz  \\
	\hhline{|b:===b:|}
	\end{tabular}
		\end{center}
\end{table}

An oversampled $\chi^{2}$ fit is used to determine the position of the
source and the flux per beam.  The best-fitting position is at
14:32:06.65$\pm$0.24 +34:16:13.4$\pm3.4$ (J2000) (quoting 3$\sigma$
position uncertainties), with a flux per beam of 23.2 $\pm$ 7.9\,mJy
(the 1$\sigma$ uncertainty comprises the background noise, primary
calibrator and gain calibration errors).  The signal-to-noise in the
detection is 3.6.  The additional photometry for SSG~1 (Table 1) 
is from the {\it Chandra} XBo\"{o}tes survey (Murray et al. 2005), NOAO-NDWFS (Jannuzi \& Dey 1999), FLAMEX (Elston et al., 
2005, in preparation), {\it Spitzer}-IRAC
Shallow survey (Eisenhardt et al.  2004), {\it Spitzer}-MIPS (IRS and MIPS GTO teams) and the Westerbork deep 1.4\,GHz survey (de Vries et al.  2002).  The upper limits are quoted to 3$\sigma$.  The corresponding multiwavelength positions and probability of coincident detection are given in Table 2.

\begin{table}
	\caption{Coincident detection positions and probability of 
accidental overlap within the 99\% confidence region of SSG~1}
	\label{tab:positions}
     \begin{center}\begin{tabular}{||cc|cc||}
	\hhline{|t:====:t|} & & & \\ [-0.15 cm]
	\multicolumn{2}{||c|}{Position (J2000)}     &     Band          &
	P(Accidental Overlap) \\
	\multicolumn{1}{||c}{RA} & \multicolumn{1}{c|}{Dec} & & \\ [0.1 cm]
	\hhline{||----||}
	\hhline{||----||} & & & \\ [-0.15 cm]
	14:32:06.61 & +34:16:12.0       &     R &  9.5\%\\
	14:32:06.55 & +34:16:11.6       &     K$_{s}$ & 2..9\%\\
	14:32:06.58 & +34:16:12.0       &     3.6\,$\mu$m &  9.4\%\\
	14:32:06.60 & +34:16:11.5       &     24\,$\mu$m &  0.2\%\\
	14:32:06.65 & +34:16:13.4       &     350\,$\mu$m & n/a\\
	\hhline{|b:====b:|}
\end{tabular}\end{center}
\end{table}

\begin{figure}[bhtp]
    \centerline{\epsfig{file=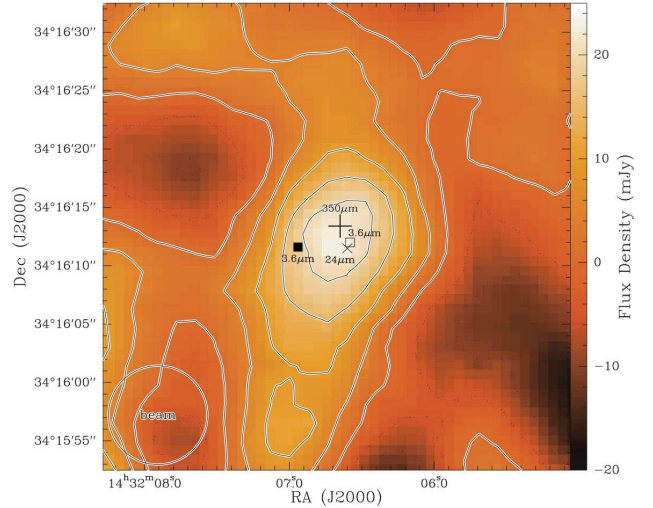,width=3.25in}}
    \caption[Submm map]{350$\,\mu$m continuum emission map of SSG~1,
    as imaged by SHARC~II. The SHARC~II (plus) and MIPS position
    (cross) are overplotted, along with the resolved IRAC 3.6\,$\mu$m
    positions of SSG~1 (open square) and SSG~1E (filled square).  The contours are in
levels of $5\,$mJy/beam.\label{submm}}
\end{figure}

Within the 99\% 350\,$\mu$m confidence region there are two candidate 
sources resolved in the optical and NIR bands (Figure 1).  We assume 
that the optical-NIR counterpart of SSG~1 is the source coincident 
with the MIPS 24\,$\mu$m detection.  The source to the east of this (at IRAC position RA$=14^h32^m6.94^s$, Dec$=+34^\circ 16'11.6''$ /J2000) is not detected by MIPS, and is designated as SSG~1E for the purposes of this Letter.

\section{CONSTRAINING THE NATURE OF SSG~1}

\begin{figure}[bhtp]
    \centerline{\epsfig{file=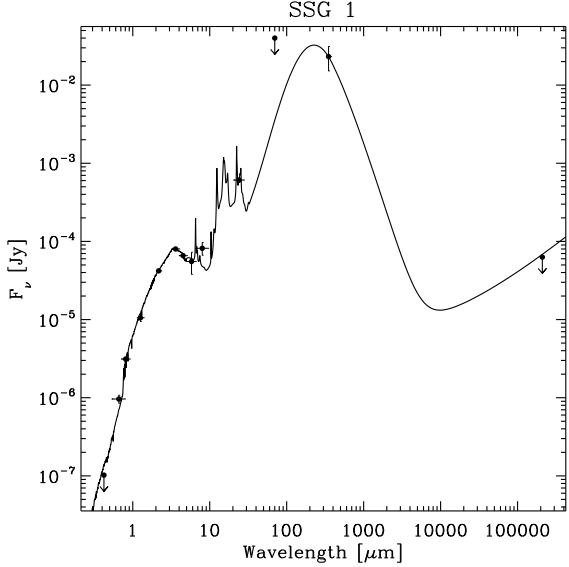,width=3.25in}}
    \caption[SED]{Best-fitting observing frame SED to SSG~1 from 
STARDUST2\label{SED}}
\end{figure}

Using the fluxes presented in Table 1, we can use
several methods to constrain the redshift, reddening, infrared
luminosity and dust temperature of SSG~1.  It can be seen in 
Figure 2 that the optical-NIR SED of SSG~1 is characterized by a prominent bump
associated with the continuum emission of the stellar populations.
This bump peaks at 1.6\,$\mu$m in the rest-frame, providing a very good
constraint on the redshift, and also indicates SSG~1 is more
likely to be starburst rather than AGN-dominated (which are usually associated with a featureless power-law spectrum; see, e.g., Egami et al.
2004).  

\begin{table}
	\caption{Best-fitting photometric redshifts and extinction for 
SSG~1 and SSG~1E with I{\scriptsize MP}Z and STARDUST2}
	\label{tab:impz}
     \begin{center}\begin{tabular}{||c|c|c|c||}
	\hhline{|t:====:t|} & & & \\ [-0.15 cm]
	Source     &     Photometric Redshift & Model    & A$\rm_{V}$ \\
	\hhline{||----||}
	\hhline{||----||} & & & \\ [-0.15 cm]
SSG~1   &  1.0$^{+0.10}_{-0.05}$ & I{\scriptsize MP}Z   & 2.6$\pm$0.2  \\
  &          0.99$^{+0.04}_{-0.03}$ & STARDUST2  & 2.2$\pm$0.09  \\
SSG~1E  &  1.0$\pm$0.05 & I{\scriptsize MP}Z & 0.5$\pm$0.1  \\
	\hhline{|b:====b:|}
\end{tabular}\end{center}
\end{table}

I{\scriptsize MP}Z (Babbedge et al. 2004) uses only the 
optical-NIR photometry when fitting templates with
Bayesian statistics.  The best-fitting
redshifts are z=1.0$^{+0.1}_{-0.05}$ for SSG~1 and z=1.0$\pm$0.05 for SSG~1E
respectively (1$\sigma$), and are listed in Table 3.  It is worth noting that if, as I{\scriptsize MP}Z suggests, both galaxies are at the same distance, and interacting, it would make the SSG pair a widely separated interacting ULIRG system (a local example of which is IRAS~09111--1007; Khan et al. 2005).

STARDUST2 (Chanial et al., 2005, in preparation) uses a mid-IR to radio
spectral library constrained by the IR-radio correlation and a variety
of local, {\it IRAS}, {\it ISO} and SCUBA color-color correlations in addition to
a stellar synthesis model for the FUV to NIR window (Devriendt et al.
1999).  This provides a simultaneous constraint on the
thermal dust emission as well as the redshift.  The effective dust temperature is found following the methodology given in Chapman et al. (2002; 2005).  The model assumes $\rm H_0=70\,km\,s^{-1}\, Mpc^{-1}$, $\rm \Omega_{m}=0.3$, and $\Omega_{\Lambda}=0.7$, and that a single component is responsible for both the optical and IR emission.  Because of the degeneracy in fitting $\rm T_{dust}$ and the Total-IR Luminosity, $\rm L_{TIR}$ (8--1100\,$\mu$m), it is the 350\,$\mu$m flux in conjunction with the radio upper limit that constrains the infrared luminosity.  The radio upper limit (Table 1) is used to find $\chi^{2}$ as follows:

$\chi^2=\sum \chi^2_i$ with the radio $\chi^2_i$ term being
\begin{eqnarray*}
\chi^2_{\rm radio } & = & [f_\nu({\rm 1.4\,GHz}) - 3\sigma]^2/\sigma^2, f_\nu({\rm 1.4\,GHz}) > 3\sigma\\
                & = & 0, \quad f_\nu({\rm 1.4\,GHz}) \leq 3\sigma
\end{eqnarray*}
where $f_\nu$ is the modeled spectral energy distribution and 3$\sigma$ 
is the radio upper limit listed in Table 1.

The best-fitting model for SSG~1 from STARDUST2 returns a redshift of 
0.99$^{+0.04}_{-0.03}$ (all errors are quoted to 1\,$\sigma$; note: STARDUST2 uses the longer wavelengths, in addition to the optical-NIR photometry, to fit the extinction given in Table 3).  This model also gives a dust temperature of 30.3$\pm$4.5\,K, and a log($\rm L_{TIR}$) of 12.02$^{+0.22}_{-0.24}$\,L$_{\odot}$, with 88\% of the total bolometric luminosity radiated at wavelengths longward of 5\,$\mu$m.  The predicted 850\,$\mu$m flux is 2.7$^{+1.7}_{-0.7}$\,mJy.  

Other redshift estimates were independently obtained following
an approach that mostly relies on the fit of the 1.6\,$\mu$m feature 
(see, e.g., Le Floc'h et al. 2004).  In agreement with the results derived
from I{\scriptsize MP}Z and STARDUST2, these fits of the stellar bump led to a
redshift z$\sim$1 using the photometric Arp220 SED, and z=1.25$\pm$0.25 using
various templates from the library of Devriendt et al. (1999).

We also use the radio-submillimeter correlation to estimate a minimum
redshift for the source, as the radio flux is unlikely to be
significantly AGN-enhanced.  Formally, these correlations should be used as a statistical redshift indicator, so with that caveat we present the minimum redshift of SSG~1 based on the 1.4\,GHz upper limit (Table 1) and an
850\,$\mu$m flux derived from the best-fitting STARDUST2 SED (Figure 2).
Using the relations of Dunne, Clements \& Eales (2000) and Carilli \&
Yun (2000a; 2000b) we get $\rm z_{min}$ of 1.2 and 1.5 respectively.  

The photometric redshift is more secure than the radio-submillimeter correlation since the latter is prone to systematics caused by uncertain dust temperatures (see Clements et al. 2004). 

\section{DISCUSSION}

The (R--$\rm K_{s})_{Vega}$ color of 5.7 for SSG~1 classifies it as an Extremely Red
Object (ERO), and both I{\scriptsize MP}Z and STARDUST2 find this object 
to be highly reddened (Table 3).  At least a third of the SMG population can be classified as EROs
(Smail et al. 2002; Webb et al. 2004; Frayer et al. 2004), which comprise two
classes of galaxies: elliptical and dusty star-forming luminous
infrared galaxies (but a significant submillimeter flux is usually
indicative of the latter class).  EROs are thought to comprise a
significant fraction of the cosmic star formation density at redshifts
of one and higher (Cimatti et al.  2002), the epoch by which the majority of
the universe's star formation has taken place (Dickinson et al. 2003; Rudnick et al. 2003).  

Determining the effective dust temperature of the source allows us to estimate the total fraction of the bolometric luminosity that is reprocessed by the dust.  Our effective dust temperature from STARDUST2 is 30.3$\pm$4.5\,K, 
which, with our total infrared luminosity of $\rm 1.0\times10^{12}\,L_{\odot}$, allows us to directly compare SSG~1 with {\it IRAS} and SCUBA-selected ULIRGs in the L vs T diagram of Blain et al.  (2004).  We find that SSG~1 is warmer
than an equivalent luminosity SMG from the Chapman et al. (2003; 2005) high redshift sample, and cooler than the {\it IRAS} moderate redshift
counterparts of Stanford et al.  (2000).  Instead, it is similar to
the intermediate redshift {\it ISO} sample of Garrett (2002).

The notion of SSG~1 being an {\it ISO} ULIRG analog is further
reinforced by the best-fitting 60-100\,$\mu$m colors from
STARDUST2, of which the closest template is FN1-40, an {\it ISO} cold ($\rm
T_{dust}$=25.7\,K) ULIRG at z=0.45 (Chapman et al.  2002).  The 
850\,$\mu$m flux ($\sim$3\,mJy) predicts SSG~1 to be fainter 
than the majority of SCUBA-selected SMGs ($\rm S_{850}\geq5\,mJy$, 
due to survey and instrument limitations).  We conclude that SSG~1 is more like {\it ISO}-selected ULIRGs rather than the SCUBA-selected SMGs.

\section{CONCLUSION}

We report the detection of the first short-wavelength
submillimeter-selected galaxy (SSG~1) and present optical, NIR and IR
photometry which we use to constrain the redshift, luminosity and dust
temperature of this source.  The photometric redshift estimators
are all in agreement, producing a best-fitting redshift of $\sim$1.
The dust temperature of 30.3$\pm$4.5\,K and luminosity of 
$\rm 1.0\times10^{12}\,L_{\odot}$ make SSG~1 an analog of
intermediate redshift {\it ISO}-selected ULIRGs rather than high redshift 
SCUBA-selected SMGs.  

If there exists a population of galaxies with properties similar to SSG~1, in 
a redshift space between {\it ISO}
and SCUBA, it may be argued that observations at 350\,$\mu$m, in
conjunction with the current {\it ISO}, {\it Spitzer} and SCUBA samples, could describe the
IR-luminous population from 0.3$<z<$4.0 (although the SCUBA-selected
galaxies will be more luminous).  Not only would this population bridge the gap in
redshift space, but also the gap in ``dust temperature space'' (and 
consequently ``dust mass space'').
SHARC~II thus complements SCUBA in revealing warmer SMGs that might be
missed otherwise.  Furthermore, finding the lower luminosity SMGs is necessary
for an accurate understanding of the submillimeter luminosity
function.

From our understanding of the first 350\,$\mu$m-selected object,
SSG~1, we infer that observations at 350\,$\mu$m, far from being
superfluous to SCUBA, will be of great use in determining the nature
and evolution of the luminous infrared galaxy population.

\section{ACKNOWLEDGEMENTS}

The Caltech Submillimeter Observatory is supported by NSF contract
AST-0229008.  This work is based in part on observations made with the
{\it Spitzer} Space Telescope, which is operated by the Jet Propulsion
Laboratory, California Institute of Technology under NASA contract
1407.  This work made use of images and data products provided by
the NOAO Deep Wide-Field Survey (Jannuzi \& Dey 1999), which is
supported by the National Optical Astronomy Observatory (NOAO).  NOAO
is operated by AURA, Inc., under a cooperative agreement with the
National Science Foundation.   The FLAMEX Survey acknowledges support from NOAO and NSF (AST-9875448,
AST-0407085, AST-0436681), and technical assistance from S.N. Raines.

We would like to thank the anonymous referee for their insightful comments which have significantly improved this Letter.  We also thank Tom Phillips and the CSO for observing time and assistance during our runs, and Darren Dowell, Colin Borys and Attila Kov\'acs for instrument and data reduction support.  We express our gratitude to our fellow GSFC co-Is on SUDS: Bob Silverberg and Dave
Chuss.  We also thank SUDS co-I Rick Arendt for continual support on the
data analysis.  S.A.K. thanks Jon Gardner, Rob Ivison and Steve Eales for very helpful discussions.


\clearpage

\clearpage

\clearpage

\clearpage

\end{document}